\documentclass[reprint,superscriptaddress]{revtex4-1}
\usepackage[utf8]{inputenc}
\usepackage{graphicx}
\usepackage{amsfonts}
\usepackage{amsmath}
\usepackage{amssymb}
\usepackage{latexsym}
\usepackage{indentfirst}
\usepackage{subfigure}
\usepackage{color}
\usepackage{mathtools}
\usepackage[colorlinks=true,urlcolor=blue,citecolor=blue]{hyperref}
\usepackage{lipsum}
\usepackage{makeidx}

\begin{document}
\title{ Extinction transitions in correlated external noise }
\author{Alexander H. O. Wada}
\affiliation{Department of Physics, Missouri University of Science and Technology, Rolla, MO 65409, USA}
\affiliation{Instituto de F\'isica, Universidade de S\~ao Paulo, Rua do Mat\~ao, 1371,\\ 05508-090 S\~ao Paulo, S\~ao Paulo, Brazil}

\author{Matthew Small}
\affiliation{Department of Physics, Missouri University of Science and Technology, Rolla, MO 65409, USA}

\author{Thomas Vojta}
\affiliation{Department of Physics, Missouri University of Science and Technology, Rolla, MO 65409, USA}
\affiliation{Kavli Institute for Theoretical Physics, University of California, Santa Barbara, CA 93106-4030, USA}

\begin{abstract}

We analyze the influence of long-range correlated (colored) external noise on
extinction phase transitions in growth and spreading processes. Uncorrelated environmental
noise (i.e., temporal disorder) was recently shown to give rise to an unusual infinite-noise
critical point [Europhys. Lett. {\bf 112}, 30002 (2015)]. It is characterized by
enormous density fluctuations that increase without limit at criticality. As a result,
a typical population decays much faster than the ensemble average which is dominated by
rare events. Using the logistic evolution equation as an example, we show here that
positively correlated (red) environmental noise further enhances these
effects. This means, the correlations accelerate the decay of a typical population but slow down the
decay of the ensemble average. Moreover, the mean time to extinction of a population
in the active, surviving phase grows slower than a power law with population size.
To determine the complete critical behavior of the extinction transition,
we establish a relation to fractional random walks, and we perform extensive
Monte-Carlo simulations.

\end{abstract}

\date{\today}

\maketitle


\section{Introduction}

Extinction transitions in spreading and growth processes have attracted
great interest in both physics and biology. From a physics point of view, extinction
transitions are prototypical nonequilibrium phase transitions between active (fluctuating)
states and inactive (absorbing) states. Nonequilibrium phase transitions display
cooperative behavior over large distances and long times analogous to phase transitions
in thermodynamic equilibrium. Understanding and classifying the resulting critical
behaviors has been a prime topic in statistical physics (for reviews, see, e.g.,
Refs.\ \cite{MarroDickman99,Hinrichsen00,Odor04,HenkelHinrichsenLuebeck_book08}).
However, growth and extinction of biological populations have been studied
scientifically for centuries, starting with the work of Euler and Bernoulli in the
18th century \cite{Euler1748,Bernoulli1760}. Whereas early approaches relied
on simple deterministic equations, recent models include fluctuations in space
and time as well as features such as heterogeneity and mobility (see, e.g.,
Refs.\ \cite{Bartlett61,Britton10,OvaskainenMeerson10}).

In the time evolution of a (model) population, several sources of noise, i.e., randomness
need to be distinguished. First, there is internal noise (also called demographic noise)
that stems from the stochastic character of the population's time evolution itself.
As this noise acts locally on each member of the population, it is averaged out for large
populations and thus vanishes in the infinite-population limit. Second, there can be
external noise, for example due to random variations of the environmental conditions
in time. In the language of statistical physics, such external noise corresponds to
temporal disorder. Spatially extended populations can also be subject to spatial disorder
which is known to have dramatic effects on nonequilibrium phase transitions
\cite{Noest86,HooyberghsIgloiVanderzande03,*Hoyos08,VojtaDickison05,*VojtaFarquharMast09,*Vojta12,VojtaLee06}
but is not considered in this paper.

External (environmental) noise plays an important role for the dynamics of populations
close to the extinction transition. In contrast to intrinsic (demographic) noise, external
noise leads to significant fluctuations of the population size with time
even for very large populations. As a result, extinction becomes easier. Specifically,
in the presence of (uncorrelated) external noise, the mean time to extinction increases only
as a power of the population size whereas it grows exponentially for time-independent environments
\cite{Leigh81,Lande93}. Effects of noise correlations (noise color) on these results
have been controversially discussed in the literature (see, e.g., Ref.\ \cite{OvaskainenMeerson10}
and references therein).

Recently, Vojta and Hoyos \cite{VojtaHoyos15,BarghathiVojtaHoyos16} demonstrated that 
external noise leads to highly unusual critical behavior, dubbed infinite-noise criticality,
at the extinction transitions of both the space-independent logistic evolution equation 
and the spatially extended contact process. 
At an infinite-noise critical point, the (relative) width of the
population density distribution
increases without limit with time, leading to enormous density
fluctuations on long time scales. This implies that the density of a typical system decays
much faster than the ensemble average of the density which is dominated by rare events.
Infinite-noise critical behavior can be seen as counterpart of infinite-randomness 
critical behavior \cite{Fisher92,*Fisher95,Vojta06} in
spatially disordered systems, but with exchanged roles of space and time.

In the present paper, we generalize the concept of infinite-noise criticality to the case
of long-range (power-law) correlated external noise. We use the logistic evolution equation
as the prime example, contrasting the cases of positively correlated (red) noise and
anticorrelated (blue) noise. We find that positively correlated (red) external
noise further enhances the infinite-noise characteristics. This means, the correlations
accelerate the decay of a typical population but slow down the
decay of the ensemble average at criticality. Moreover, the mean time to extinction of a population
in the active, surviving phase grows slower than a power law with population size.
Our paper is organized as follows. In Sec.\ \ref{SEC:LE}, we introduce the logistic
evolution equation with external noise (temporal disorder), and we describe the mapping
onto a reflected random walk. We also summarize the results for uncorrelated disorder.
In Sec.\ \ref{SEC:theory}, we develop the theory of infinite-noise criticality for
power-law correlated noise. Sec. \ref{SEC:MC} is devoted to extensive Monte-Carlo simulations.
We conclude in Sec.\ \ref{SEC:conclusions}.

\section{Logistic equation \label{SEC:LE}}

\subsection{Definition \label{SEC:LE:Def}}

The logistic equation is a prototypical model for growth and
spreading processes in nature \cite{Verhulst}.
It reads
\begin{equation} \label{EQ:LE}
	\frac{d\rho}{d t} = -\mu \rho + \lambda \rho (1-\rho),
\end{equation}
where $\rho$, $\mu$ and $\lambda$ are the density of individuals,
the death rate and the growth rate, respectively.
If $\mu$ and $\lambda$ are time independent constants, the logistic equation can be solved exactly,
\begin{equation} \label{EQ:LE_clean}
\rho^{-1} = \rho^{-1}_0 e^{(\mu-\lambda)t} + \lambda \frac{e^{(\mu - \lambda)t} -1}{\mu-\lambda},
\end{equation}
where $\rho_0$ is the initial density at time $t=0$.
In this solution, the population survives if $\lambda > \mu$ and
dies out exponentially if $\lambda < \mu$.
Between survival and extinction there is a critical point, at
$\mu = \lambda$, where $\rho$ decays as a power law, $\rho \sim t^{-1}$.

Temporal disorder
can be introduced into the logistic equation by adding time-dependent noise $\Delta \mu$  and $\Delta \lambda$ to
$\mu$ and $\lambda$.
We consider noise that is constant during time intervals $\Delta t_n$;
our rates will thus have constant values $\mu_n = \mu + \Delta \mu_n$ and $\lambda_n = \lambda + \Delta \lambda_n$.
The noise terms have zero mean, $[\Delta \lambda_n] = [\Delta \mu_n] = 0$,
where $[\ldots]$ denotes average over disorder configurations.

Correlations between different time intervals
can be characterized by the correlation function
\begin{equation}
G_\phi(i, j) = [\phi_i \phi_j] - [\phi_i][\phi_j],
\end{equation}
where $\phi$ refers to $\Delta \lambda$ or $\Delta \mu$.

\begin{figure}
\centering
\includegraphics[scale=1.0]{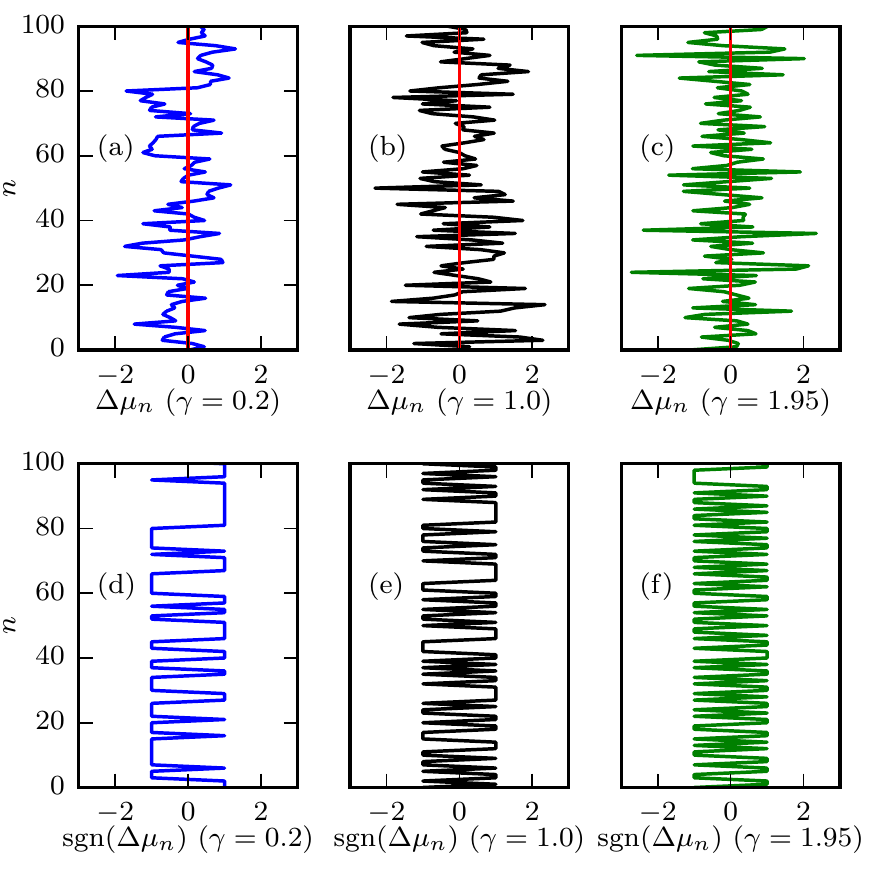}
\caption{Panels (a), (b) and (c) show examples of the fractional Gaussian noise $\Delta \mu_n$ for
         the superdiffusive ($\gamma < 1$), uncorrelated ($\gamma = 1$) and subdiffusive ($\gamma > 1$) regimes, respectively.
         To emphasize the differences, we show $\text{sgn} (\Delta \mu_n)$ 
         in panels (d), (e) and (f).
         These plots show that positive correlations ($\gamma < 1$) make the noise less likely to alternate between positive and negative values,
         while negative correlations ($\gamma > 1$) cause the noise to alternate more frequently.
          \label{EQ:CorrelatedNoise}}
\end{figure}

In the present paper, we are particularly interested in long-range
correlations that asymptotically
decay as
\begin{equation} \label{EQ:CorrFunc_asymp}
G(i, j) \sim |i - j|^{-\gamma},
\end{equation}
with $0 < \gamma < 2$.
Most of our simulations will make use of the 
fractional Gaussian noise \cite{Qian2003}
that has the correlation function
\begin{equation}\label{EQ:FBM_Corr}
	G_{\text{FBM}}(i, j)= \frac{\sigma^2}{2}[ |t+1|^{2-\gamma} -2|t|^{2-\gamma} +|t-1|^{2-\gamma} ] \underset{t \to \infty}{\sim} t^{-\gamma},
\end{equation}
with $t = j-i$.
The exponent $\gamma$ (related to the Hurst exponent by $H = (2-\gamma)/2$)
controls the color of the noise:
it is red when $\gamma < 1$ (positive correlations),
blue if $\gamma > 1$ (negative correlations),
and white at $\gamma = 1$ (uncorrelated).
Fig.\ \ref{EQ:CorrelatedNoise} shows examples of the fractional
Gaussian noise for different values of $\gamma$.

Alternatively, one could also use a generic power-law decay such as
\begin{equation} \label{EQ:Generic_Corr}
	G = \frac{1}{(1+t^2)^{\gamma/2}}.
\end{equation}
For $\gamma < 1$, this generic correlation function produces red noise just like the fractional Gaussian noise
(\ref{EQ:FBM_Corr}) because its Fourier transform $\tilde{G}(\omega)$ diverges for $\omega	\rightarrow 0$.
However, for $\gamma > 1$, $\tilde{G}(\omega)$ approaches a constant for $\omega \rightarrow 0$, i.e., the noise has a white component.
This will become important in section \ref{SEC:Harris_criterion} when we discuss the Harris criterion.

%

\subsection{Mapping to random walk \label{SEC:MRW}}

\begin{figure}
\centering
\includegraphics[scale=1.0]{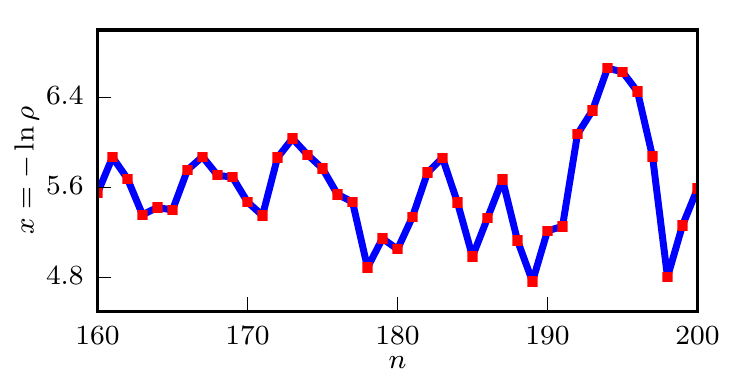}
\caption{ Density $\rho(t)$ for an individual noise configuration, plotted as
          $x = -\ln \rho$ vs.\ time interval $n$.
          The initial condition is $\rho_0 = 0.5$; the parameters are
          $\Delta t = 2$, $\lambda_n = 1$, and $\mu_n$
          is a uncorrelated Gaussian random variable with average and variance both equal to $1$.\label{FIG:RandomWalk}}
\end{figure}

The logistic equation can be solved exactly within each of the time intervals $\Delta t_n$,
during which the rates are constant.
Using the exact solution given in (\ref{EQ:LE_clean}),
we find the recursion relation
\begin{equation} \label{EQ:LE_iter}
	\rho^{-1}_{n+1} = \rho^{-1}_n a_n + c_n,
\end{equation}
for the density $\rho_n$ at the beginning of time interval $\Delta t_n$.
Here, $a_n = \exp \left\{ (\mu_n-\lambda_n)\Delta t_n \right\}$ and
$c_n = (a_n-1)\lambda_n/(\mu_n - \lambda_n)$.

Neglecting $c_n$ for the moment, and defining $x_n = -\ln \rho_n$,
this recursion can be mapped onto a random walk
$x_{n+1} = x_n + \ln a_n$.
The average $[\ln a_n]  = [ \Delta t_n(\mu_n - \lambda_n) ]$
controls the bias of the walker, whose velocity is given by
\begin{equation} \label{EQ:speed}
	v = (\mu - \lambda).
\end{equation}
The velocity is directly related to the distance
to criticality.
When $v > 0$, $x = - \ln \rho$ increases, therefore $\rho$ decreases exponentially.
Thus $v > 0$ represents the inactive phase.
The exact opposite happens for $v < 0$, thus $v<0$ is the active phase.
The critical point is located at $v = 0$.
The effect of $c_n$ can be approximated by a reflecting wall
at $x = 0$; $c_n$ is only important for larger densities and
prevents the density to rise above $\rho = 1$.

The mapping onto a random walk is illustrated in Fig.\ \ref{FIG:RandomWalk}.
As we can see, $x$ changes randomly with
increasing time.
While $x$ changes linearly at each time interval,
$\rho$ goes through an exponential growth or decay.

\subsection{Uncorrelated temporal disorder}

In Ref.\ \cite{VojtaHoyos15}, Vojta and Hoyos studied the effects of uncorrelated temporal disorder on the logistic equation
treating the external noise $\Delta \mu_n$ and $\Delta \lambda_n$
as uncorrelated random variables.
Here, we briefly  summarize  their results for later comparison with the correlated case.

In the active phase and close to criticality, 
the average stationary density scales as $\rho_\text{st} = [e^{-x}] \sim |v|$,
while the typical density behaves as $\ln \rho_{\text{typ}} =  -[x] \sim 1/v$.
The difference in the typical and average values stems from
a singularity in the probability distribution $P(\rho)$:
while $\rho_\text{st}$ is dominated by rare events at $x = -\ln \rho = 0$,
the leading contribution for $\rho_{\text{typ}}$ comes
from the singularity at small $\rho$.

At the critical point, $P(\rho)$ broadens without limit with increasing time,
justifying the term infinite-noise criticality.
The average and typical densities show different time dependencies.
The average density decays as a power law $[\rho] \sim t^{-1/2}$,
in contrast to the exponential decay for the typical density,
$- \ln \rho_{\text{typ}} \sim t^{1/2}$.

In the inactive phase, the typical and the average
densities both decay exponentially.
Close to the critical point,
$- \ln [\rho] \sim v^2 t$
and $- \ln \rho_\text{typ} \sim v t$.
Moreover, the correlation time scales as
$\xi_t \sim |v|^{-2}$.

The results for the average density and the correlation time
can be compared with our expectations for critical phenomena.
In the active phase $\rho_\text{st} \sim |v|^{\beta^{\text{unc}}}$,
at the phase transition $[\rho] \sim t^{-\delta^{\text{unc}}}$,
and in the inactive phase $\xi_t \sim |v|^{-\nu^{\text{unc}}_\parallel}$.
This comparison yields the exponents
\begin{equation} \label{EQ:Clean_LE_Exponents}
\beta^{\text{unc}} = 1, \quad \delta^{\text{unc}} = 1/2 \quad \text{and} \quad \nu^{\text{unc}}_\parallel = 2,
\end{equation}
for the (uncorrelated) infinite-noise critical point.
They are different from the exponents of the clean logistic equation
($\beta^{\text{clean}} = 1, \quad \delta^{\text{clean}} = 1 \quad \text{and} \quad \nu^{\text{clean}}_\parallel = 1$).

The lifetime $\tau$ of a population is the average time it takes to reach extinction.
Starting from $\rho_0 > 0$, the logistic equation never truly reaches extinction ($\rho = 0$).
Nonetheless, for a population of size $N$,
$\tau$ can be estimated as the time it takes $\rho$ to fall below $1/N$.
Through this definition, it was found that $\tau$ grows as
$\tau \sim N^{a|v|}$ ($a$ is a constant) in the active phase,
$\tau \sim \ln^2 N$ at criticality,
and $\tau \sim \ln N$ in the inactive phase.
These results are in agreement with the notion of a temporal Griffiths phase
caused by temporal disorder \citep{TemporalGP_Vaszquez}.


\section{Logistic equation with correlated temporal disorder: theory \label{SEC:theory}}

\subsection{ Generalized Harris criterion \label{SEC:Harris_criterion}}

Before developing our theory,
it is interesting to know if power-law correlated temporal disorder is
expected to change the critical behavior.
This can be ascertained via a generalization of the Harris criterion \cite{Harris_Criterion}.
Following \cite{VojtaDickman_HarrisCriterion},
a critical point should be stable against power-law correlated temporal disorder,
if the correlation time exponent $\nu_\parallel$ fulfills $\gamma \nu_\parallel > 2$.

Starting from the clean logistic equation ($\nu^{\text{clean}}_\parallel = 1$),
we see that any value of $\gamma < 2$ violates this criterion.
We therefore expect the clean critical behavior to be destabilized for all our values of $\gamma$.

On the other hand, we could also ask if weak correlated disorder
changes the critical behavior of the logistic equation with uncorrelated temporal disorder.
Applying the same criterion with $\nu^{\text{unc}}_\parallel = 2$,
we see that the power-law correlations should only change the critical behavior if $\gamma < 1$.

\subsection{Reflected fractional Brownian motion}

In Sec. \ref{SEC:MRW}, we have seen that the time evolution of the logistic equation with temporal disorder can be mapped onto a random walk with a reflecting wall.
In our case of long-range correlated disorder, the result of this mapping is a reflected fractional random walk (i.e., reflected fractional Brownian motion).
Reflected fractional Brownian motion is an interesting problem itself; it was recently studied via large-scale Monte-Carlo simulations \citep{Vojta_ReflectedFBM}.
Here, we briefly summarize the key results.


The probability distribution $P(x,t)$ of the position $x$ of an unbiased walker at time $t$
(if it started at  $x=0$ at $t=0$)
takes the scaling form 
\begin{equation} \label{EQ:Histogram_Colapse}
P(x, t) = \frac{1}{\sigma t^{(2-\gamma)/2}} Y\left( \frac{x}{\sigma t^{(2-\gamma)/2}} \right),
\end{equation}
where $\sigma$ is the width of the noise distribution.
From this equation, we can see that
the width of $P$ increases as $t^{(2-\gamma)/2}$.
As the walker is restricted to the positive x-axis, its average position
behaves accordingly
\begin{equation} \label{EQ:AnomalousDiffusion}
	[x] \sim t^{(2-\gamma)/2}.
\end{equation}
This means the motion is
superdiffusive for positive correlations ($\gamma < 1$),
while subdiffusion occurs when the correlations are negative ($\gamma > 1$).

However, a surprising feature of reflected fractional Brownian motion is
the non-Gaussian character of its probability distribution.
$P(x, t)$ was found to be Gaussian in $x$ for large $x$ only.
It features a power-law singularity close to the wall (for $x \rightarrow 0$)
\begin{equation} \label{EQ:PowerLaw_singularity}
Y(y) \sim y^{2/(2-\gamma) -2} \quad \text{or} \quad P(x, t) \sim \frac{x^{2/(2-\gamma)-2}}{t^{\gamma/2}}.
\end{equation}
This means $P(x, t)$ diverges for $x \rightarrow 0$ in the superdiffusive case,
while it goes to zero in the subdiffusive case.

\subsection{Critical behavior}

The results of the last subsection allow us to determine the critical behavior of the logistic equation
with long-range correlated external noise.

From the anomalous diffusion law (\ref{EQ:AnomalousDiffusion})
and the singularity (\ref{EQ:PowerLaw_singularity}) close to the wall,
we can evaluate the typical and average densities.
The leading contribution to the average density $[\rho]$
is given by the singularity, therefore
\begin{equation}  \label{EQ:rho_time_critical}
[\rho] = \int dx P(x,t) e^{-x} \sim t^{-\gamma/2}.
\end{equation}
In contrast, the typical density $\rho_\text{typ} = e^{-[x]}$
decays much faster
\begin{equation} \label{EQ:rho_typ_critical}
	\rho_{\text{typ}} = e^{- \int dx P(x,t) x } \sim e^{-Dt^{\phi}},
\end{equation}
where $D$ is a constant and $\phi = (2-\gamma)/2$.

The typical time for $\rho$ to fall below $1/N$
is also the time it takes for $x$ to reach $\ln N$.
From eq.\ (\ref{EQ:AnomalousDiffusion}), we therefore obtain the lifetime $\tau$ of a population of finite-size $N$ at criticality as
\begin{equation} \label{EQ:tau_lnN_critical}
	\tau \sim (\ln N)^{\eta},
\end{equation}
with $\eta = 2/(2-\gamma)$.

To derive the correlation time $\xi_t$,
we analyze the time it takes for the displacement of the walker to significantly deviate from its critical counterpart.
In the inactive phase, the walker simply walks to the right, far away from wall,
meaning that the movement is ballistic, with speed $v = \mu - \lambda$ as in eq.\ (\ref{EQ:speed}).
This ballistic motion overcomes the anomalous diffusion (\ref{EQ:AnomalousDiffusion})
when $|v|t \sim t^{1-\gamma/2}$, therefore the correlation time follows
\begin{equation} \label{EQ:CorrTime_mu_critical}
	\xi_t \sim |v|^{-2/\gamma}.
\end{equation}

The stationary density $\rho_\text{st}$ in the active phase can be estimated from the density reached at time $\xi_t$.
Combining (\ref{EQ:rho_time_critical}) and (\ref{EQ:CorrTime_mu_critical}) we obtain
\begin{equation} \label{EQ:rho_statCrit}
	\rho_{\text{st}} \sim \xi_t^{-\gamma/2} \sim |v|.
\end{equation}

The critical exponents can be easily extracted
from equations (\ref{EQ:rho_time_critical}),
(\ref{EQ:CorrTime_mu_critical}) and (\ref{EQ:rho_statCrit}),
giving us
\begin{equation}
\delta = \gamma/2, \quad \nu_\parallel = 2/\gamma \quad \text{and} \quad \beta = 1.
\end{equation}
%

The theory therefore predicts that the critical exponents (except $\beta$) differ from the clean ones,
in agreement with the generalized Harris criterion.
Setting $\gamma = 1$ (uncorrelated or white noise),
our predictions agree with the results reported for
the logistic equation with uncorrelated temporal disorder \cite{VojtaHoyos15} (see eq.\ (\ref{EQ:Clean_LE_Exponents})).
Furthermore, our exponents obey the scaling relation
$\delta = \beta/\nu_\parallel$.

The logistic equation does not contain a notion of space.
However, imagining a $d$-dimensional space of $N$ individuals,
we can define a length scale $L \sim N^{1/d}$.
According to (\ref{EQ:tau_lnN_critical}), this length scale behaves
as $\ln L \sim \tau^{(2-\gamma)/2}$ suggesting an exponential relation between correlation length and time
\begin{equation} \label{EQ:Temporal_ActivatedScaling}
	\ln \xi \sim \xi^{(2-\gamma)/2}_t.
\end{equation}
%
This emphasizes the infinite-noise character of the critical point,
since the logarithmic relation (\ref{EQ:Temporal_ActivatedScaling}) can be understood as the temporal
counterpart of activated scaling in spatial disordered systems \cite{HooyberghsIgloiVanderzande03, Fisher92, Vojta06}.
The exponent $\omega = (2-\gamma)/2$ is the temporal equivalent of the
tunneling exponent $\psi$.

\subsection{Temporal Griffiths phase}

(Spatial) Griffiths phases in spreading and growth processes
exist due rare spatial regions that are locally active while the bulk system
is inside the inactive phase \cite{Noest86,Vojta06}.
In contrast, temporal Griffiths phases are observed in the active phase of temporally disordered systems \cite{TemporalGP_Vaszquez}.
Here, the rare regions are long time intervals that are locally on the inactive side of the extinction transition.
Specifically, we are looking for rare time intervals in the active phase
(having effective death rate above the critical point) 
that are strong enough to make the density fall below $1/N$ (causing extinction of a population of size $N$).

A time interval (rare region) of size $T_{RR}$ has an effective death rate given by
\begin{equation}
	\mu_{RR} = \frac{1}{T_{RR}} \sum_{i \in RR} \mu_i.
\end{equation}
For extinction to occur its contribution to the density evolution should make $\rho$
fall below $1/N$
\begin{equation}
	a_{i+1} a_{i+2} \ldots a_{i+T_{RR}} = e^{(\lambda-\mu_{RR})T_{RR}} \sim 1/N.
\end{equation}
Thus the rare region size necessary for the population to become extinct is $T_{RR} \sim (\ln N)/(\mu_{RR} - \lambda)$.

Analogously to Ref.\ \cite{Ibrahim_longrangeDisCP}, the probability to find a rare region of size $T_{RR}$ and effective
death rate $\mu_{RR}$ is
\begin{equation}
	P \sim \exp \left\{ - A T^{\gamma}_{RR}  \right\},
\end{equation}
where $A$ is a constant.
This means, for $\gamma < 1$, long rare regions are more likely than in the uncorrelated case.
If we combine this equation with the rare region size,
the lifetime $\tau$, which is proportional to the inverse extinction probability,
scales as
\begin{equation} \label{EQ:Temporal_GP}
	\tau \sim \exp \left\{ C (\ln N)^\gamma \right\},
\end{equation}
Therefore, the lifetime increases more slowly than a power law in $N$ for $\gamma < 1$.
The constant $C = C(\mu)$ plays the role of the dynamical exponent
of the temporal Griffiths phase.
The dependence of $C$ on $\mu$ does not directly follow from the above argument, but it can be found from scaling.

Given the simplicity of the logistic equation,
there are only three relevant variables, time, distance from criticality
and the (artificially introduced) system size $N$,
defined as a cutoff for the density, $\rho = 1/N$
(if $\rho$ falls below $1/N$, a population of size $N$ goes extinct).
From the lifetime at criticality (\ref{EQ:tau_lnN_critical}) and $(\ln N)^{-\gamma} \sim |v|^{2-\gamma}$
(which can be derived from eq.\ (\ref{EQ:rho_statCrit}) and (\ref{EQ:Temporal_ActivatedScaling})),
we can infer the scaling function
\begin{equation} \label{EQ:tau_scaling}
	\tau = (\ln N)^{2/(2-\gamma)} \Phi ( |v|^{(2-\gamma)} (\ln N)^{\gamma} ).
\end{equation}

Comparing (\ref{EQ:Temporal_GP}) and (\ref{EQ:tau_scaling}),
we get $\Phi(z) \sim \exp(C' z)$ ($C'$ is a constant),
therefore the lifetime in the temporal Griffiths phase should, more precisely, behave as
\begin{equation} \label{EQ:Temporal_GP_withprefactor}
	\tau \sim \exp \left\{ C' |v|^{2-\gamma} (\ln N)^\gamma \right\}.
\end{equation}
%
Moreover, setting $\gamma = 1$, we obtain the power law $\tau \sim N^{C'|\mu - \mu_c|}$,
just as in the uncorrelated case.

\section{Monte-Carlo simulations \label{SEC:MC}}

\subsection{Overview}

To test the predictions resulting from the mapping to the fractional random walk, we now perform
Monte-Carlo simulations by numerically iterating the recursion relation (\ref{EQ:LE_iter}).
The relevant simulation parameters are:
the average rates $\mu$ and $\lambda$,
the noise $\Delta \mu_n$ and $\Delta \lambda_n$,
and $\Delta t$.

The recursion (\ref{EQ:LE_iter}) is invariant under the simultaneous
transformation of the rates and the time interval according to
$\lambda_n \rightarrow M \lambda_n$, $\mu_n \rightarrow M \mu_n$, $\Delta t \rightarrow M^{-1} \Delta t$.
We can use this freedom to choose numerically convenient parameters.

The average rates control the distance to criticality.
We fix $\lambda = 128$ and tune the extinction transition by varying $\mu$.
The extinction transition occurs at $\mu = \mu_c = \lambda = 128$.
$v = \mu - \mu_c > 0$ means we are in the inactive phase
(bias of the walker is to the right).
In contrast, $v = \mu - \mu_c < 0$ implies a bias towards the wall (located at $x = 0$),
thus the system is in the active phase.


Concerning the temporal disorder, we set $\Delta  \lambda = 0$.
The noise terms $\Delta \mu _n$ of the death rates are Gaussian distributed
numbers following the long-range correlation function (\ref{EQ:FBM_Corr}).

The Gaussian correlated random numbers
are generated by the Fourier filtering method \cite{Stanley96_CorrRNG}.
Given the uncorrelated Gaussian random numbers $\nu_n$ and
the correlation function $G(n)$,
we calculate their Fourier transforms
$\tilde{\nu}_n$ and $\tilde{G}(n)$.
The correlated random numbers $\mu_n$ are given by the inverse Fourier
transform of
\begin{equation}
	\tilde{\mu}_n = \tilde{\nu}_n \left\{\tilde{G}(n)\right\}^{1/2}.
\end{equation}



All simulations start from $\rho = 1$ ($x= 0$), and run up to
$2^{26}$ ($\approx 6.7 \times 10^7$) iterations.
Unless otherwise mentioned, we set the variance of $\Delta \mu_n$ to  $\sigma^2 = 1$ and $\Delta t = 2$.
With these parameters, the typical change of the density $\rho$ over one time interval at criticality is
approximately $e^{\pm \sigma \Delta t} = e^{\pm 2}$.





\subsection{Critical behavior}

\begin{figure}
\centering
\includegraphics[scale=1.0]{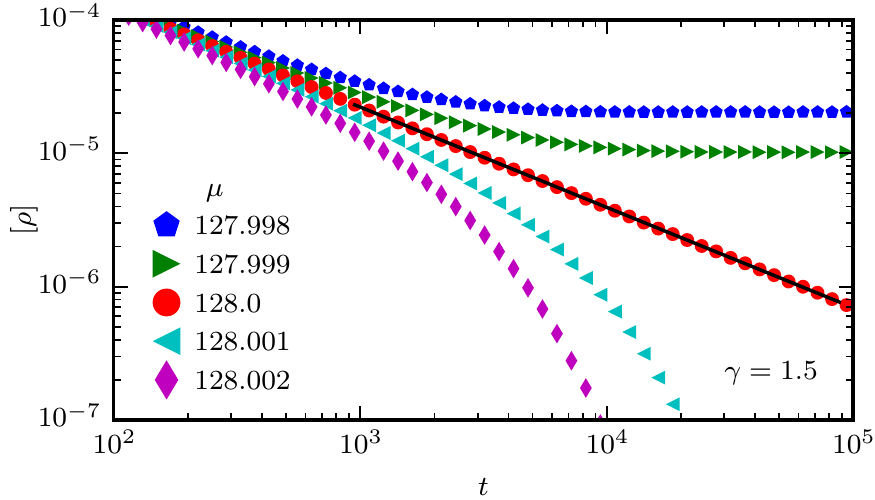}
\caption{ Average density $\rho$ as a function of $t$ close to the critical point for $\gamma = 1.5$.
        Below $\mu = \mu_c = 128$ the population survives by reaching a stationary density whereas
        it decays if $\mu> \mu_c$.
        Between survival and extinction there is a critical point at $\mu = \mu_c$.
        The solid black line shows a power-law fit whose
        exponent $0.75(1)$ agrees with the one predicted by eq. (\ref{EQ:rho_time_critical}).
        The data are averages over $10^7$ disorder configurations. \label{FIG:DensityNearCriticalPoint}}
\end{figure}

\begin{figure}
\centering
\includegraphics[scale=1.0]{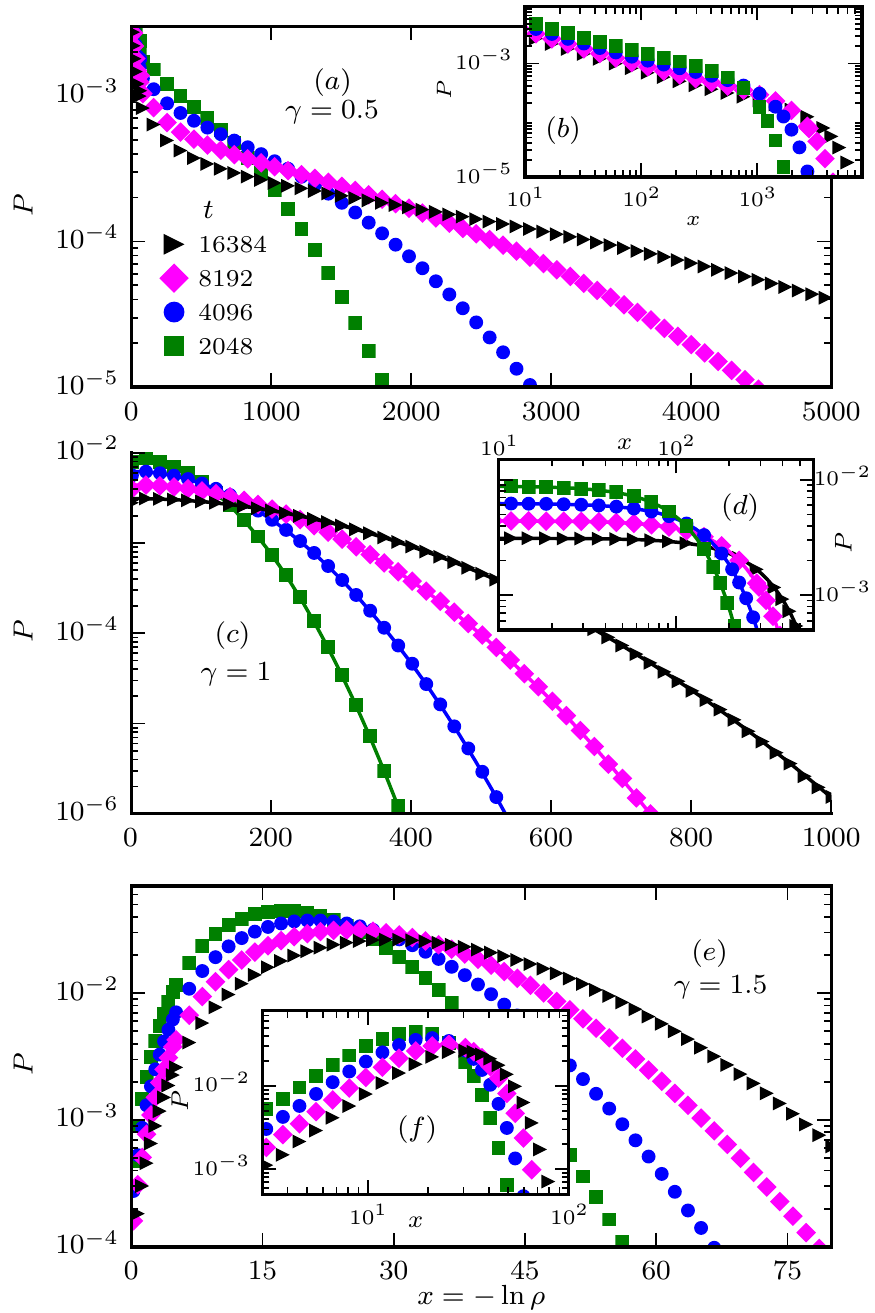}
\caption{ Probability distribution $P$ as function of $x = -\ln \rho$ at criticality and at different times $t$,
        using $\sigma = 1$ and $\Delta t = 2$.
        Panel (a) shows the singularity at $x= 0$ for $\gamma = 0.5$  (red noise or superdiffusive regime).
        In the double-logarithmic plot in panel (b), we can see that the singularity takes a power-law form.
        Panel (c) shows the uncorrelated case, $\gamma = 1$, where $P$ follows the half Gaussian $2 \exp \left\{ x^2/(\sigma^2 \Delta t^2 t) \right\}/\sqrt{2 \pi \sigma \Delta t^2 t}$
        (represented by the solid lines).
        Panel (d) displays this same data, but in a double-logarithmic scale.
        Panel (e) shows that the singularity is also present for $\gamma = 1.5$ (subdiffusive regime or blue noise),
        however, $P$ gets depleted instead of diverging.
        Panel (f) displays the same data as in panel (e), however in a double-logarithmic plot,
        again demonstrating the power-law behavior of the singularity.
        These simulations run up to $t = 2^{16}$ ($\approx 16 \times 10 ^3$) and average 
        over $10^7$ disorder realizations.
        Uncertainties are smaller than the symbol sizes. \label{FIG:Histograms} }
\end{figure}

\begin{figure}
\centering
\includegraphics[scale=1.0]{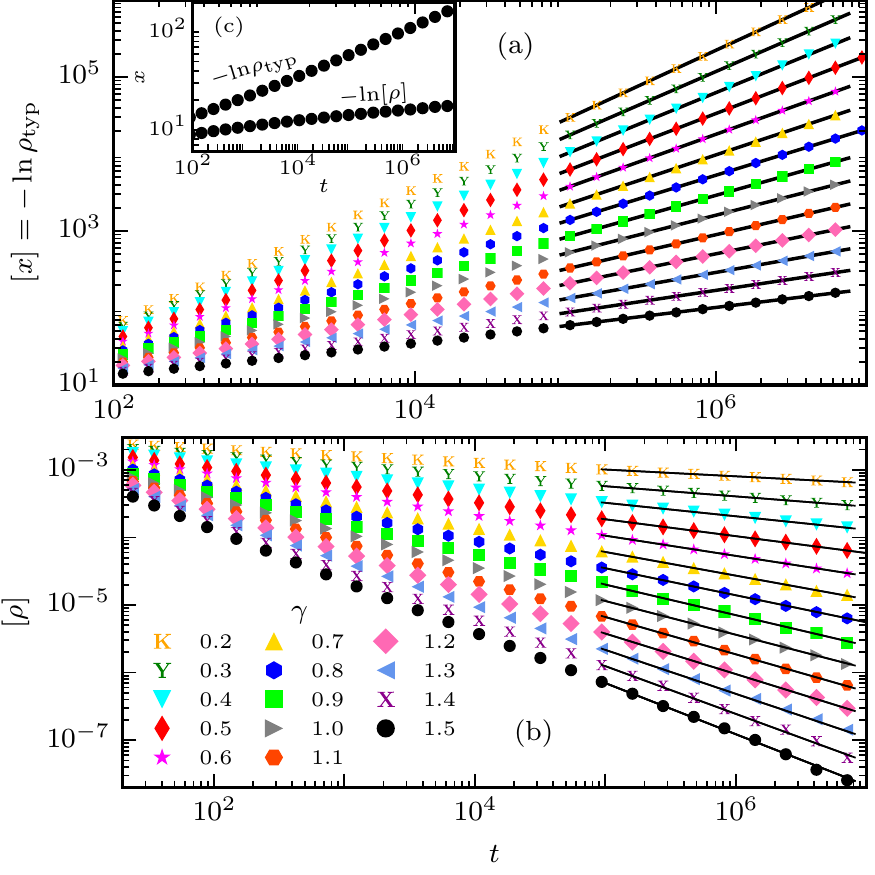}
\caption{ (a) Typical density at criticality plotted as $[x] = -[\ln \rho] = -\ln \rho_\text{typ}$ vs.\ time $t$ for several $\gamma$.
          (b) Average density $[\rho]$ vs.\ time $t$.
          In both cases the curves follow power laws for more than two decades in time for all values of $\gamma$ we have simulated.
          The inset, panel (c), compares $- \ln \rho_\text{typ} $ and $- \ln [\rho]$,
          for $\gamma = 1.5$.
          Each curve averages over $10^6$ disorder configurations,
          making uncertainties about the size of the symbols.
          \label{FIG:AnomalousDiff}}
\end{figure}

\begin{figure}[h!]
\centering
\includegraphics[scale=1.0]{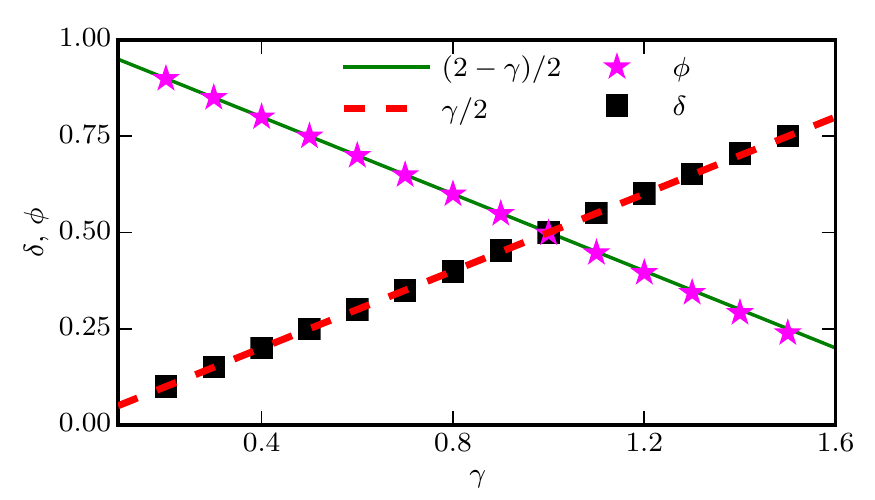}
\caption{ Exponents $\delta$ and $\phi$ extracted from the data in Fig.\ \ref{FIG:AnomalousDiff}.
        The solid and dashed (green and red) lines are our theoretical predictions $\phi = (2-\gamma)/2$ and $\delta = \gamma/2$.
        All exponents agree with the predictions within the uncertainty bars that are smaller than the symbol size.
        The small deviations in the exponent $\phi$ for $\gamma \geq 1.3$
        are also observed in fractional Brownian motion \cite{Vojta_ReflectedFBM} and stem from corrections to scaling. \label{FIG:DeltaAnon}}
\end{figure}

We first perform simulations near the critical point $\mu_c = \lambda = 128$
to check if the phase transition still exists in the presence of correlations.
Fig.\ \ref{FIG:DensityNearCriticalPoint} shows the average density
$[\rho]$ as a function of time for $\gamma = 1.5$.
This plot verifies that the extinction transition exists at
the expected $\mu_c$; for $\mu > \mu_c$ the population decays,
while for $\mu < \mu_c$ it has a non-zero stationary density.
Other values of $\gamma$ leads to the same qualitative behavior near $\mu_c$.

Now we turn our attention to the critical probability distribution $P(x,t)$
of the logarithmic density $x=-\ln\rho$ at time $t$.
Fig.\ \ref{FIG:Histograms} displays the probability distribution
as a function of $x$ for several values of the correlation exponent $\gamma$.

In all cases, the probability distribution broadens without limit for $t \rightarrow \infty$,
demonstrating the infinite-noise character of the critical point.
Moreover, Fig.\ \ref{FIG:Histograms}{\color{red}{a}} shows that $P(x, t)$ develops a divergence for $x \rightarrow 0$
in the case of positive (persistent) correlations ($\gamma = 0.5$).
The same data is plotted in a double-logarithmic scale in panel (b),
revealing this singularity to be a power law.
Panels (c) and (d) display simulations for the uncorrelated case ($\gamma = 1$) together with
the half Gaussian, $P(x, t) = 2\exp \{ -x^2/(2\sigma^2\Delta t^2 t) \}/\sqrt{2 \pi \sigma^2 \Delta t^2 t}$,
expected from the drift-diffusion equation with flux-free boundary condition at $x=0$.
Our simulation results for the uncorrelated case match very well with the half Gaussian.
The power-law singularity is also verified in the case
of negative correlations ($\gamma = 1.5$),
as shown in panels (e) and (f).
In this case, $P$ gets depleted close to $x = 0$.
These results are in very good agreement with the predictions of the reflected random walk theory.

After analyzing the probability distribution, we look for
important implications: do average and typical density
scale differently?
Fig.\ \ref{FIG:AnomalousDiff} shows our results
for $[x] = - \ln \rho_{\text{typ}}$ (panel a) and
$[\rho]$ (panel b) as functions of time,
with $\gamma$ ranging from $0.2$ to $1.5$.


The data in both panels are compatible with the expected power laws
over two decades in time ($- \ln \rho_\text{typ} = [x] \sim t^{\phi}$
and $[\rho] \sim t^{-\delta}$).
We can thus extract their exponents to compare them with the values
predicted by (\ref{EQ:rho_time_critical}) and (\ref{EQ:rho_typ_critical}).
Fig.\ \ref{FIG:DeltaAnon} shows that the extracted exponents
match our predictions, $\phi = (2-\gamma)/2$ and $\delta = \gamma/2$, very well,
although there are small deviations in
the exponent $\phi$ due to strong corrections to scaling
deep in the subdiffusive regime ($\gamma$ approaches 2).

\begin{figure}
\centering
\includegraphics[scale=1.0]{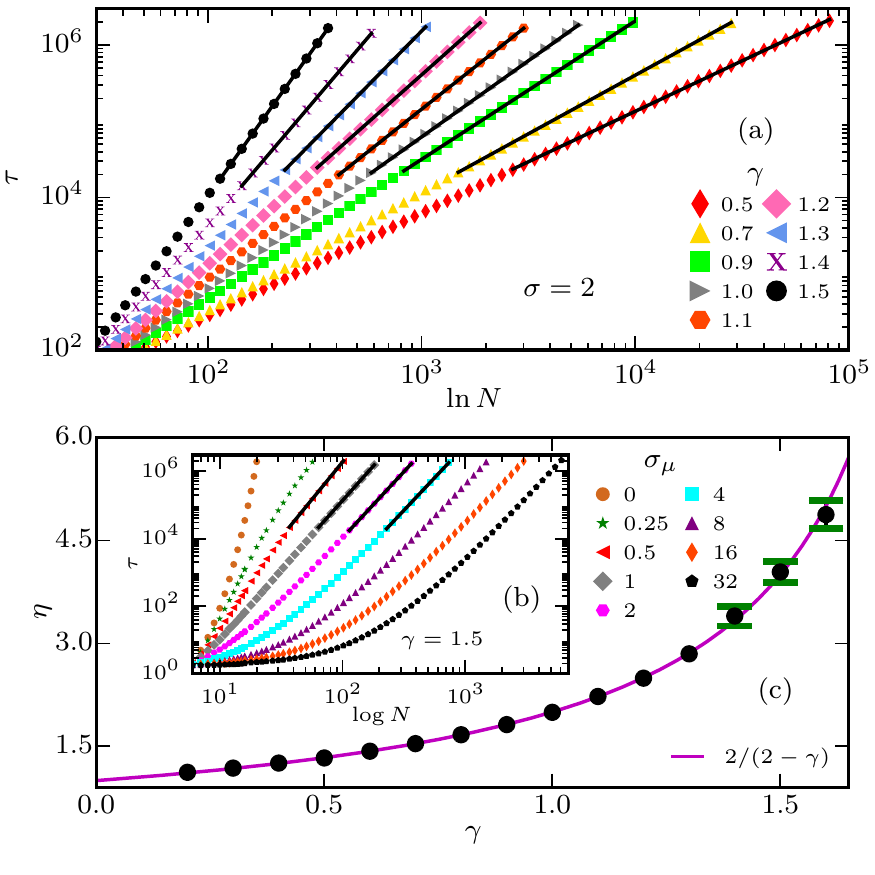}
\caption{(a) Lifetime $\tau$ at the critical point as a function of the population size $N$ for several $\gamma$ using $\sigma = 2$.
         The expected behavior (\ref{EQ:tau_lnN_critical}) corresponds to a straight line.
         (b) Effect of the disorder strength $\sigma$ on the lifetime for $\gamma = 1.5$.
         (c) Exponents were determined by fitting the lifetime to $\tau \sim (\ln N)^{\eta}$.
         The uncertainties are smaller than the symbol size unless explicitly shown.
         The solid magenta line represents the prediction $\eta = 2/(2-\gamma)$.
         Each curve is averaged over a number of disorder configurations that ranges from $10^{4}$ to $10^5$.
          \label{FIG:CriticalTau}}
\end{figure}

To finish our analysis of the properties at criticality, $\mu = \mu_c$,
we run simulations for the lifetime $\tau$ of a population of size $N$.
In Fig.\ \ref{FIG:CriticalTau}{\color{red} a}
we plot $\tau$ vs.\ $\ln N$ for several
values of $\gamma$.
Our data follows (\ref{EQ:tau_lnN_critical}) for almost
two decades in $\tau$, therefore we can proceed to extract the exponent.
When performing the fits, we realized that changing the value of 
$\sigma$ significantly changes the crossover to the asymptotic behavior.
The crossover for small $\sigma$ can be understood as the crossover
from the clean logistic equation to the temporally disordered one,
since $\sigma = 0$ represents absence of temporal disorder.
This implies that for small $\sigma$, the system will follow the clean behavior at short times before crossing over to the asymptotic regime.
This is illustrated in Fig \ref{FIG:CriticalTau}{\color{red} b}.
For large $\sigma$, the asymptotic regime is only reached for large $N$
because small populations can go extinct in a single time step.


To determine the optimal $\sigma$ that leads to the fastest crossover,
we took only fits with reduced $\chi^2 < 1.5$.
Although the range of optimal $\sigma$ seems to change slightly with $\gamma$,
values of $\sigma$ between $0.5$ and $4$ generally yield good results.


Repeating the analysis for several $\gamma$, we found the exponents
shown in panel (c).
The figure demonstrates that our measured exponents match very well
with the prediction (\ref{EQ:tau_lnN_critical}) $\eta = 2/(2-\gamma$).

\subsection{Off-critical properties}

\begin{figure}
\centering
\includegraphics[scale=1.0]{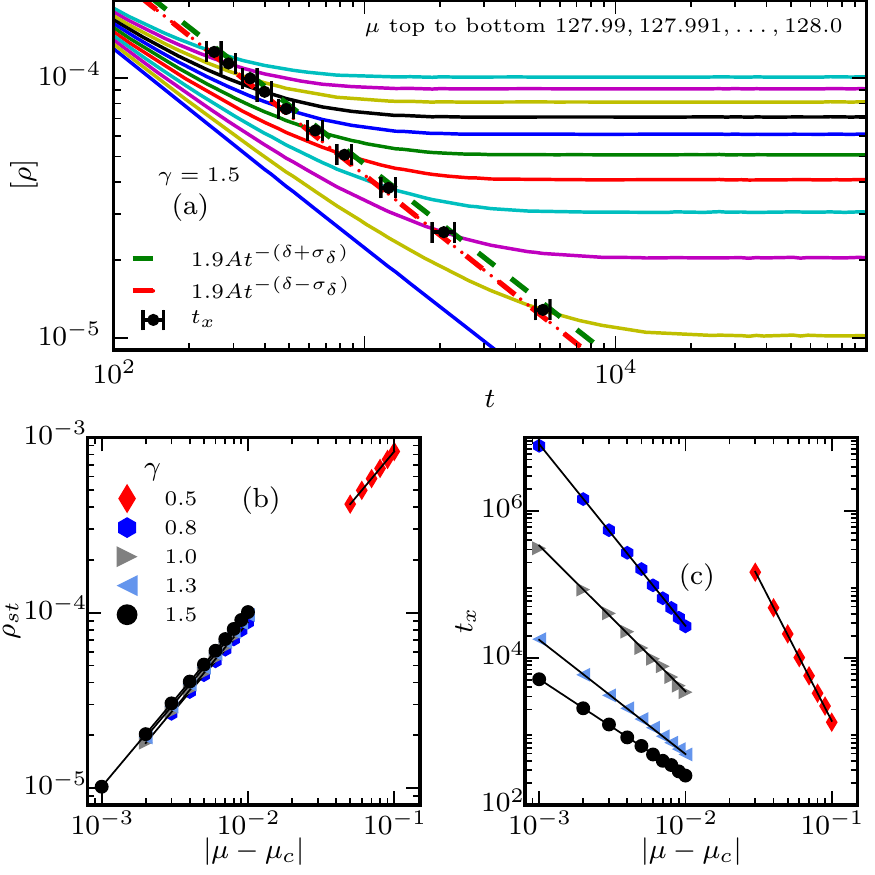}
\caption{ Off-critical simulations.
         Panel (a) displays $[\rho]$ as a function of time for $\gamma = 1.5$ and several $\mu$.
         The stationary density $\rho_\text{st}$ is obtained fitting a constant to the long time behavior of $[\rho]$.
		 Representing the critical curve as $At^{-\delta}$, the dashed (green) and dash-dotted (red) lines show $1.9 \rho_\text{crit}$ with one uncertainty in $\delta$. The crossover time $t_x$ (black circles) is evaluated as the average time between those curves and its uncertainty is the half width.
         In panels (b) and (c) we can see that $\rho_\text{st}$ and $t_x$ follow power laws in $|\mu - \mu_c|$ for several $\gamma$.
         In panel (b), uncertainties are much smaller than the symbol size.
         Uncertainties in panel (c) are of the order of the symbol size.
         \label{FIG:OffCriticalDensity}}
\end{figure}

In the last section we found the numerical results at criticality to be
in perfect agreement with the predictions of the reflected fractional Brownian motion theory.
Specifically, the exponent $\delta$ agrees with the prediction $\delta = \gamma/2$.
To complete the set of critical exponents, we analyze
the stationary density $\rho_\text{st}$ and
the correlation time $\xi_t$ 
to obtain $\beta$ and $\nu_\parallel$.

\begin{figure}
\centering
\includegraphics[scale=1.0]{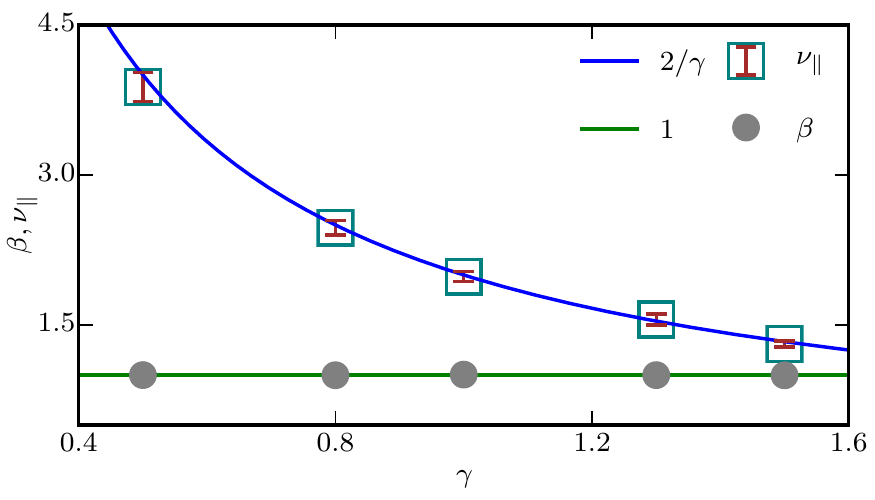}
\caption{ Critical exponents $\beta$ and $\nu_\parallel$ extracted from the data in Fig.\ \ref{FIG:OffCriticalDensity}.
		  Simulations for $\gamma < 1$ required times up to $2^{26} \approx 6.7 \times 10^7$ due the long crossover to the asymptotic behavior.
          All values agree with the theoretical predictions.
          The uncertainty in $\beta$ is of the order of $10^{-2}$. 
           \label{FIG:BetaNu}}
\end{figure}

Panel (a) of Fig.\ \ref{FIG:OffCriticalDensity} shows simulations
close to the critical point.
When $\mu < \mu_c$, $[\rho]$ reaches
the stationary density $\rho_\text{st}$ in the long-time limit.
The stationary density is evaluated by fitting a constant to the density data at long times
and then plotted in panel (b) vs.\ $|\mu-\mu_c|$.
As all values of $\rho_\text{st}$ are very close to each other
for different values of $\gamma$,
correlations have only a small impact on the stationary density.
The behavior of $\rho_\text{st}$ vs.\ $|\mu-\mu_c|$ is compatible with power laws and
yields exponents that agree with $\beta = 1$, as is shown
in Fig.\ \ref{FIG:BetaNu}.

As an estimate of the correlation time $\xi_t$,
we take the time $t_x$ at which the off-critical
density equates $1.9$ times the critical one.
The power-law behavior of $t_x$ vs.\ $|\mu - \mu_c|$ can be
observed in Fig.\ \ref{FIG:OffCriticalDensity}{\color{red}c},
and the extracted exponents $\nu_\parallel$ are shown in Fig.\ \ref{FIG:BetaNu}.
The exponents agree very well with the relation $\nu_\parallel = 2/\gamma$ predicted in eq.\ (\ref{EQ:CorrTime_mu_critical}).

\subsection{Temporal Griffiths phase}

\begin{figure}
\centering
\includegraphics[scale=1.0]{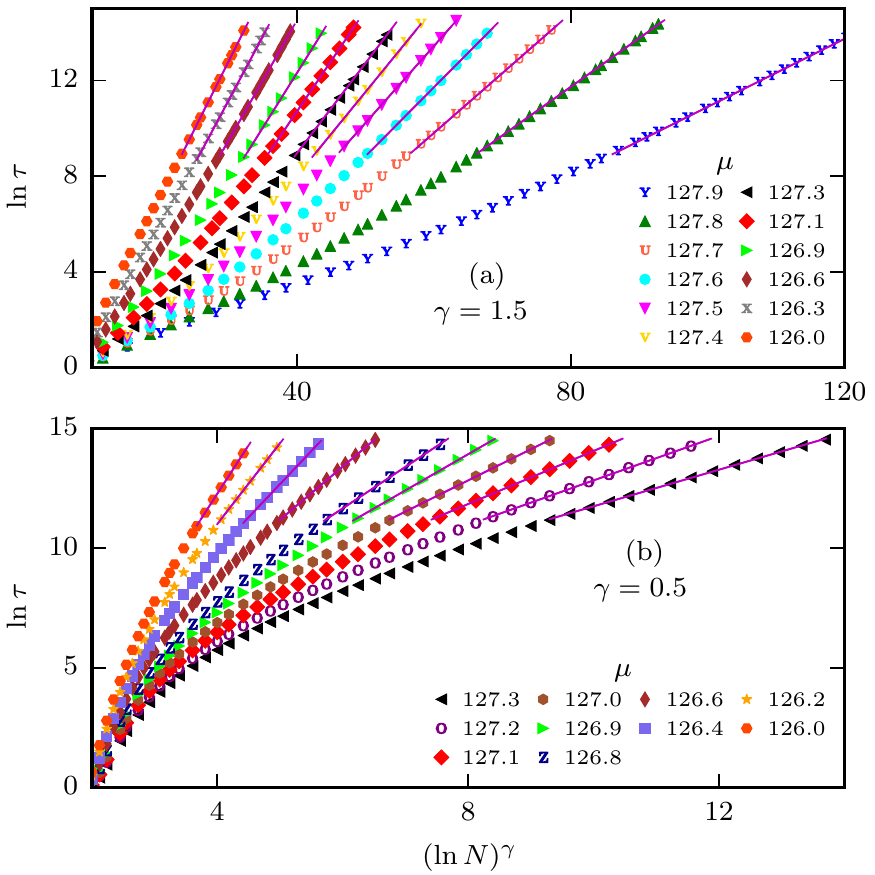}
\caption{ Lifetime $\tau$  vs.\ population size $N$ plotted as $\ln \tau$ vs.\ $(\ln N)^\gamma$,
          such that straight lines indicates a behavior compatible with eq.\ (\ref{EQ:Temporal_GP}).
          For $\gamma = 1.5$ (panel (a)), straight lines can be fitted over almost three decades in time and for $\mu$ within $2\sigma$ of $\mu_c$ ($\sigma = 1$).
          However, for $\gamma = 0.5$, we see a smaller range in time and $\mu$.
          The prefactor of $[\ln N] ^\gamma$ in eq.\ (\ref{EQ:Temporal_GP}) was also evaluated and plotted in Fig.\ (\ref{FIG:GP_Prefactor}).  \label{FIG:TemporalGP}}
\end{figure}

\begin{figure}
\centering
\includegraphics[scale=1.0]{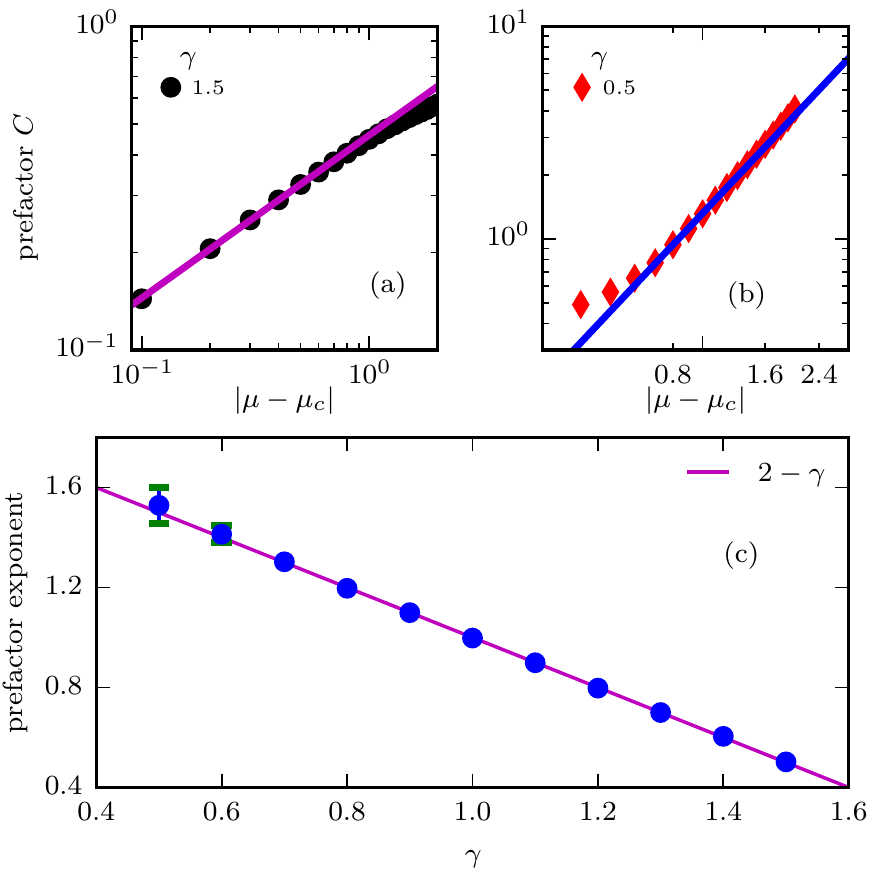}
\caption{ Analysis of the prefactor $C$ of equation (\ref{EQ:Temporal_GP}), $\tau \sim \exp \left\{ C(\ln N)^\gamma \right\}$.
          Panels (a) and (b) shows $C$ extracted from the data in Fig.\ \ref{FIG:TemporalGP} vs.\ $|\mu - \mu_c|$.
          The prefactor $C$ is expected to behave as a power law $C = C'|\mu - \mu_c|^{2-\gamma}$ (see eq.\ (\ref{EQ:Temporal_GP_withprefactor})).
          Panel (c) shows the prefactor exponent extracted from the data (blue circles) and our prediction $2-\gamma$ (solid magenta line).
          Uncertainties are of the order of the symbol size if not shown.\label{FIG:GP_Prefactor}}
\end{figure}

So far, all our simulations results are in agreement
with the reflected fractional Brownian motion theory.
What about the temporal Griffiths phase?

To observe the temporal Griffiths phase, we run simulations on the
active side of the extinction transition ($\mu < \mu_c$), but with $\mu$ close enough to $\mu_c$,
such that there are extended time intervals with $\mu_n > \mu_c$.
Specifically, we run simulations up to $2 \sigma$ away
from the critical point
for a minimum chance of approximately $2.5\%$
to get a value of $\mu_n$ in the inactive phase.

Fig.\ \ref{FIG:TemporalGP} shows
the lifetime as a function of the population size $N$ plotted as $\ln \tau$ vs.\ $(\ln N )^\gamma$,
such that the predicted behavior (\ref{EQ:Temporal_GP_withprefactor}) will look like a straight line.
For $\gamma = 1.5$ (panel (a)), our data for large $N$ follow straight
lines for almost three decades in $\tau$.
In contrast, for $\gamma = 0.5$, our plot shows a very
long crossover to the behavior predicted by eq.\ (\ref{EQ:Temporal_GP_withprefactor}).
This is illustrated in panel (b),
where we can see straight lines
for only about one and half decades in $\tau$.
The range of $\mu$ is also significantly smaller:
while the data for $\gamma = 1.5$ display the expected behavior
for $\mu$ up to $2\sigma$ away from $\mu_c$,
for $\gamma = 0.5$ the predicted behavior has a smaller range.

The slow crossover can be understood as follows.
Positive correlations ($\gamma < 1$) enhance the fluctuations,
therefore it takes longer for the bias $v = \mu - \mu_c$ (\ref{EQ:speed}) to take over.
This means that all off-critical properties,
such as the stationary density, the crossover time $t_x$ and
the temporal Griffiths behavior (Figs.
\ref{FIG:OffCriticalDensity}{\color{red}b}, \ref{FIG:OffCriticalDensity}{\color{red}c} and
\ref{FIG:TemporalGP}),
require longer simulation times to reach their true asymptotic behaviors.
The exact opposite happens in the presence of negative correllations ($\gamma > 1$).

To further verify our theory, we can use the lifetime data
to find the prefactor multiplying
$[\ln N]^\gamma$ in eq.\ (\ref{EQ:Temporal_GP_withprefactor})
and compare it to $C'(\mu-\mu_c)^{2-\gamma}$.
Panels (a) and (b) of Fig.\ \ref{FIG:GP_Prefactor}
show the prefactor extracted from Fig.\ \ref{FIG:TemporalGP} for $\gamma = 1.5$ and $\gamma = 0.5$, respectively.
The data follow power-law behavior in both cases.
Deviations for small $|\mu - \mu_c|$ (for $\gamma = 0.5$) can be attributed
to the slow crossover from the critical regime to the asymptotic behavior.
Too far away from $\mu_c$, rare regions becomes rarer,
thus longer times are again required to achieve the true asymptotic behavior.

\begin{figure}
\centering
\includegraphics[scale=1.0]{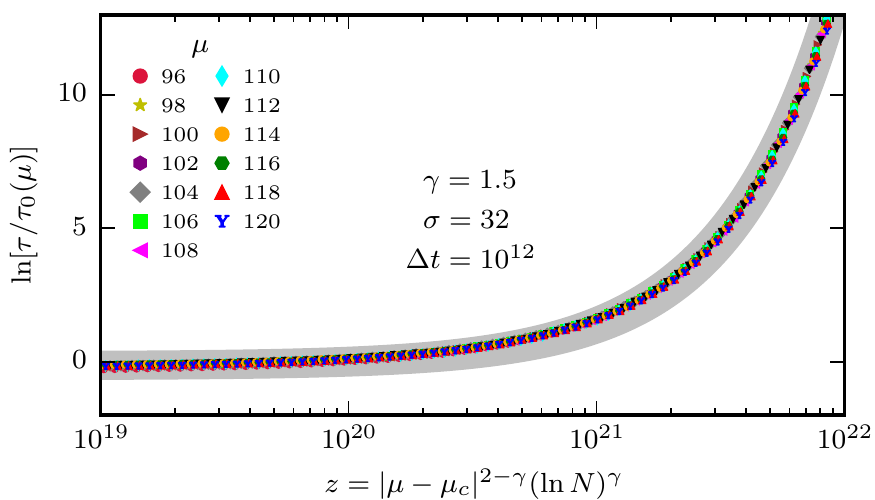}
\caption{Scaling collapse of the lifetime in the temporal Griffiths phase.
         To achieve this collapse we employed very high (unphysical) disorder $\sigma = 32$ and $\Delta t = 10^{12}$.
         The correction to scaling $\tau_0(\mu)$ was obtained adjusting $\tau_0(\mu) = B(\mu) \exp \left\{ C [\ln N]^\gamma \right\} $
         in every curve.
         The grey background is a plot of $\tau = D_1 e^{D_2 z}$ with the appropriate parameters.\label{FIG:Colapse}}
\end{figure}

Following this argument, we discarded too small and large values of $|\mu - \mu_c|$.
The exponent of the prefactor,
shown in panel (c) of Fig.\ \ref{FIG:GP_Prefactor},
follows the theoretical prediction $2-\gamma$ for $\gamma$ between $0.5$ and $1.5$.

Even though the prefactor has the expected behavior,
we could not obtain a perfect scaling collapse
as implied by (\ref{EQ:tau_scaling}).
A nearly perfect scaling collapse can be obtained for very strong disorder
$\sigma = 32$ and $\Delta t = 10^{12}$
\footnote{
It is worthwhile noting that this strong disorder
is unphysical, because at each step the change
in $\rho$ should be of the order of
$a_n = \exp \left\{ \pm \sigma \Delta t \right\} = \exp \left\{ \pm 32\times10^{12} \right \}$,
which is clearly unrealistic for any population.
Moreover, since we are doing simulations near $\mu_c =128$,
setting $\sigma = 32$ put us only $4\sigma$ away from $\mu < 0$.
This allows negative $\mu$ (negative death rates) to appear in our simulations;
they could be interpreted as a spontaneous creation of organisms
},
as shown in Fig.\ \ref{FIG:Colapse}.
This plot uses the fuctional form
$\tau = \tau_0(\mu) \exp \left\{ C'(\mu - \mu_c)^{2-\gamma} [\ln N]^\gamma \right\}$,
where $\tau_0(\mu)$ acts as a correction to scaling.

All the above results were for the active phase.
On the inactive side of the transition,
the walker simply walks ballistically to the right,
far away from the wall,
regardless of its correlated steps.
This way, we expect the lifetime to behave as
in the uncorrelated case $\tau \sim \ln N$.
We have numerically verified this behavior
for $\gamma = 0.5$ and $1.5$.

\section{Conclusion \label{SEC:conclusions} }

In summary, we have studied the effects of power-law correlated temporal disorder on the extinction transition of the logistic equation.
Temporal disorder (external noise) was introduced by making 
the death rate a random function of time,
whose correlations decay as $t^{-\gamma}$.

The generalized Harris criterion predicts the critical point of the logistic equation
to be destabilized in the presence of correlated temporal disorder.
Two cases must be distinguished.
First, we apply the generalized Harris criterion the clean logistic equation.
In this case, the criterion predicts the temporal disorder to
destabilize the clean critical behavior for all $\gamma < 2$.
Second, we apply the generalized Harris criterion to the logistic equation
with uncorrelated temporal disorder.
In this case, the correlations are only expected to be relevant if $\gamma < 1$.
In other words, \emph{if the system already contains uncorrelated (white) noise},
additional correlated noise will only change the behavior if the correlations fulfill $\gamma < 1$.

To develop a theory of the extinction transition,
we have mapped our problem onto fractional Brownian motion with a reflecting wall.
At the critical point,
the reflected fractional random walk theory predicts that
the probability distribution $P(x, t)$ of the logarithmic density $x = -\ln \rho$
broadens without limit with increasing time,
in perfect agreement with the notion of infinite-noise criticality \cite{VojtaHoyos15}.
Moreover, $P(x,t)$ features a power-law singularity close to $x = 0$.
As a result, the average density decays as a power law in time (\ref{EQ:rho_time_critical}),
while the typical density decays exponentially (\ref{EQ:rho_typ_critical}).
The lifetime of a finite-size population increases as a power of the logarithm of its size (\ref{EQ:tau_lnN_critical}).
Compared to the uncorrelated case \cite{VojtaHoyos15},
positive (negative) power-law correlations, $\gamma < 1$ ($\gamma > 1$),
results in a slower (faster) decay of the average density,
a faster (slower) decay of the typical density,
and a slower (faster) growth of the lifetime.

In the active phase,
the lifetime grows more slowly then exponential with time,
in perfect agreement with the notion of temporal Griffiths phases
caused by temporal disorder \cite{TemporalGP_Vaszquez}.
More specifically, the lifetime of the population
grows slower (faster) than a power of the population size (\ref{EQ:Temporal_GP_withprefactor})
if the correlations are positive (negative).

All these predictions were confirmed by large scale Monte-Carlo
simulations, for times up to $2^{26} \approx 6.7 \times 10^7$.

Over the course of this paper,
we have shown results for the fractional Gaussian noise
with the correlation function (\ref{EQ:FBM_Corr}).
Does a generic power-law correlation function,
such as (\ref{EQ:Generic_Corr}),
affect the critical point in the same way?
As explained in Sec. \ref{SEC:LE:Def},
the generic correlation function (\ref{EQ:Generic_Corr})
has a white (uncorrelated) component,
therefore we expect the correlations to be relevant
only for $\gamma < 1$.
In this case, the critical behavior will be identical to that produced by the fractional Gaussian noise (\ref{EQ:FBM_Corr}).
If $\gamma > 1$, in contrast, the white component overwhelms the
blue (anticorrelated) noise, hence the behavior is the
same as for the uncorrelated case ($\gamma = 1$).
We have numerically verified this behavior
by performing simulations using (\ref{EQ:Generic_Corr})
for $\gamma = 1.5$, $0.8$, $0.6$, and $0.4$.
These results are in agreement with the generalized Harris criterion discussed above.

It is interesting to compare the infinite-noise critical behavior
induced by long-range correlated temporal disorder with the infinite-randomness behavior
which is induced by long-range spatial disorder in the contact process \cite{Ibrahim_longrangeDisCP}.
For $\gamma < 1$ and close to the critical point,
the correlation length $\xi^{\text{CP}}$ of the
1d contact process with quenched correlated spatial disorder \cite{Ibrahim_longrangeDisCP}
scales as $\xi^\text{CP} \sim r^{-2/\gamma}$, where $r$ is the distance from the critical point.
At the critical point, $r = 0$, we have activated scaling $(\xi^\text{CP})^{(2-\gamma)/2} \sim \ln \xi^\text{CP}_t$.
In the inactive phase, we observe the (spatial) Griffiths phase behavior $\rho \sim \exp \left\{ - C^\text{CP} r^{2-\gamma} (\ln t)^\gamma \right\}$,
where $C^\text{CP}$ is a constant.
These relations are the spatial counterparts of equations (\ref{EQ:CorrTime_mu_critical}),
(\ref{EQ:Temporal_ActivatedScaling}) and (\ref{EQ:Temporal_GP_withprefactor}).

What about critical behavior of the stationary density (\ref{EQ:rho_statCrit})?
As shown by Rieger and Iglói in Refs.\ \cite{Rieger_PRB1998, Rieger_PRL1999},
the average surface magnetization $[m_s]$ of the random transverse-field Ising model
can be mapped onto a random walk.
This mapping allow us to compare the critical behavior of $[m_s]$ with $\rho_\text{st}$.
For $\gamma < 1$, Rieger and Iglói found that $[m_s]$ varies linearly with distance from criticality,
in agreement with eq.\ (\ref{EQ:rho_statCrit}).

How generic is the infinite-noise critical behavior
observed in the logistic equation with correlated temporal disorder?
As the disorder only affects the linear growth and death rates, we expect our results to hold for all single-variable growth models
with disorder in the linear rates and
a non-linear term restricting the population growth.
In contrast,
if the leading growth and/or decay terms in the evolution equation are nonlinear in $\rho$,
the time evolution of $\rho$ in each of the time intervals will not be exponential,
necessitating modifications to our theory.

In this work we have studied a model devoid of spatial structure.
Nonetheless, spatial structure is also an important factor
in the population dynamics. 
Therefore we are currently investigating the effects of power-law
correlated temporal disorder on the contact process by means of Monte-Carlo
simulations.

Extinction transitions (i.e. absorbing state phase transitions) have been observed in systems
ranging from liquid crystals \cite{Chate_PRL2007} and driven suspensions \cite{LCorte_ROPDS2008,Franceschini_PRL2011}
to bacteria colonies \cite{KSKorolev_CC2011, KSKorolev_PG2011}.
Our theory describes the effects of external noise on such transitions.
They could be studied, for example, by growing yeast or bacteria in well-mixed media under fluctuating environmental conditions.

\section*{Acknowledgements}

This work was supported in part by the NSF under Grant Nos.\ PHY-1125915
and DMR-1506152 and by the São Paulo Research Foundation (FAPESP) under
Grant No.\ 2017/08631-0. T.V.\ is grateful for the hospitality of the Kavli
Institute for Theoretical Physics, Santa Barbara where part of the work
was performed.

\bibliographystyle{apsrev4-1}
\bibliography{Paper.bib}

\end{document}